# Mammographic density: Comparison of visual assessment with fully automatic calculation on a multivendor dataset

**Daniela Sacchetto[1], Lia Morra[1], Silvano Agliozzo[1], Daniela Bernardi[2], Tomas Bjorklund[3], Beniamino Brancato[4], Patrizia Bravetti[5], Luca A. Carbonaro[6], Loredana Correale[1], Carmen Fantò[2], Elisabetta Favettini[7], Laura Martincich[8], Luisella Milanesio[9], Sara Mombelloni[10], Francesco Monetti[11], Doralba Morrone[4], Marco Pellegrini[2], Barbara Pesce[12], Antonella Petrillo[13], Gianni Saguatti[14], Carmen Stevanin[15], Rubina M. Trimboli[6], Paola Tuttobene[2], Marvi Valentini[2], Vincenzo Marra[16], Alfonso Frigerio[9], Alberto Bert[1], Francesco Sardanelli[6,17]**

[1] *im3D S.p.A., Research and Development Dept., Turin, Italy*
[2] *APSS, Trento, Italy*
[3] *KTH, Technology and Health, Stockholm, Sweden*
[4] *ISPO, Florence, Italy*
[5] *Centro prevenzione oncologica, Ravenna, Italy*
[6] *IRCCS Policlinico San Donato, Milan, Italy*
[7] *ASLBi, Biella, Italy*
[8] *IRCC, Candiolo, Italy*
[9] *Regional Reference Centre for Breast Cancer Screening, Turin, Italy*
[10] *Ospedale Valduce, Como, Italy*
[11] *Ospedale S. Martino, Genova, Italy*
[12] *C.d.C. Paideia, Rome, Italy*
[13] *IRCCS, National Cancer Institute "G. Pascale" Foundation of Naples, Naples, Italy*
[14] *AUSL, Bologna, Italy*
[15] *Ospedale Regionale, Bolzano, Italy*
[16] *OIRM S.Anna, Turin, Italy*
[17] *Università degli Studi di Milano, Department of Biomedical Sciences for Health, Milan, Italy*

**Abstract**

**Objectives:** To compare breast density (BD) assessment provided by an automated BD evaluator (ABDE) with that provided by a panel of experienced breast radiologists, on a multivendor dataset.

**Methods:** Twenty-one radiologists assessed 613 screening/diagnostic digital mammograms from 9 centers and 6 different vendors, using the BI-RADS *a*, *b*, *c*, and *d* density classification. The same mammograms were also evaluated by an ABDE providing the ratio between fibroglandular and total breast area on a continuous scale and, automatically, the BI-RADS score. Panel majority report (PMR) was used as reference standard. Agreement (κ) and accuracy (proportion of cases correctly classified) were calculated for binary (BI-RADS *a-b* versus *c-d*) and 4-class classification.

**Results:** While the agreement of individual radiologists with PMR ranged from κ=0.483 to κ=0.885, the ABDE correctly classified 563/613 mammograms (92%). A substantial agreement for binary classification was found for individual reader pairs (κ=0.620, standard deviation [SD]=0.140), individual versus PMR (κ=0.736, SD=0.117), and individual versus ABDE (κ=0.674, SD=0.095). Agreement between ABDE and PMR was almost perfect (κ=0.831).

**Conclusions:** The ABDE showed an almost perfect agreement with a 21-radiologist panel in binary BD classification on a multivendor dataset, earning a chance as a reproducible alternative to visual evaluation.

**Key points:**

- Individual BD assessment differs from PMR with κ as low as 0.483.

- An ABDE correctly classified 92% of mammograms with almost perfect agreement (κ=0.831).

- An ABDE can be a valid alternative to subjective BD assessment.

**Abbreviations and acronyms:** ABDE (automated breast density evaluator); BD (breast density); PMR (panel majority report); SD (standard deviation).

## Introduction

Breast density (BD) is an important parameter in breast imaging. Higher BD is believed to be associated with higher breast cancer risk [1], albeit its exact role and the magnitude of its independent effect is still controversial [2]. Importantly, sensitivity of mammography is significantly reduced in women with higher BD [3]; data from screening programs show123 that interval cancers are more frequent in women with dense breasts [4].

To increase effectiveness of screening programs, personalized screening strategies taking into account individual risk are drawing increasing attention. Strategies for tailoring screening according to BD have been experimented and are ongoing, including additional imaging to mammography, such as ultrasound, breast digital tomosynthesis, or magnetic resonance imaging [5, 6, 7, 8], or reducing the screening interval for women with high BD [9].

Most available data on BD impact on breast cancer risk and sensitivity of mammography are based on visual assessment according to the scoring system introduced by the American College of Radiology in the context of the Breast Imaging Reporting and Data System (BI-RADS) [10], commonly used in clinical practice [11]. BI-RADS classifies BD into four classes, according to the relative amount of fibroglandular tissue: *a*, almost entirely fat; *b*, scattered fibroglandular; *c*, heterogeneously dense; and *d*, extremely dense. However, visual assessment is subjective and associated with suboptimal reproducibility. Several studies have investigated inter- and intra-observer variability of visual BI-RADS density classification, finding only moderate or substantial agreement [11,12,13]. As a consequence, the replacement of visual assessment by an automated reproducible classification has111213 been advocated [13].

Quantitative BD assessment on a continuous scale, as provided by an automated system, could also be used to track density changes over time [1] and to shed more light on the controversial role of BD for breast cancer risk and tailoring screening programs. However, relatively few data are available on the correlation between automated quantitative BD estimates and visual BI-RADS evaluation, causing interpretation issues of the automated results to radiologists. A study [11] has recently explored the correlation between density automated measure and BI-RADS visual scoring: a correspondence between the two was suggested, but only mammograms b11y one vendor were included in the dataset.

Our aim was to test a new automated BD evaluator (ABDE) for mammographic density assessment on a multivendor dataset of digital mammograms and to compare its results with the BI-RADS scores provided by a large panel of experienced breast radiologists.

**Materials and methods**

Im3D S.p.A. (Torino, Italy) provided technical support for the study. Three authors (XX, YY and ZZ) are researchers at im3D, and two authors (JJ and KK) are consultants for im3D. Non-consultant authors had full control of the data and information submitted for publication.

Institutional Review Board approval and written informed consent was waived because the study retrospectively evaluated a dataset of fully anonymized images acquired within routine diagnostic procedures.

*Study dataset*

A set of 664 digital mammograms was retrospectively randomly collected from nine centers in Italy. Exams presenting surgical scars, substantial technical defects (e.g. large skin wrinkles due to breast compression), and evident lesions to visual inspection to one experienced breast radiologist (LAC) were excluded. A final dataset of 613 exams was obtained: 377 (61.5%) were "diagnostic" examinations (including women self-referring to mammography for subjective symptoms, follow-up, or spontaneous screening) while 236 (38.5%) came from organized population-based screening programs.

Images were acquired using digital mammography equipment from six vendors: Amulet FDR 1000 (Fujifilm Corporation, Minato-ku, Tokyo, Japan), Senograph DS version ADS_53.40 (General Electric Healthcare, Little Chalfont, Bucks, UK), Selenia Dimensions (Hologic, Bedford, MA, USA), Giotto Image 3DL and TOMO (Internazionale Medico Scientifica, Bologna, Italy), MicroDose Mammography (MDM) L30 (Sectra AB, Linköping, Sweden) and Mammomat Inspiration (Siemens, Munich, Germany). The dataset vendor distribution is reported in Figure 1.

Average age was 55 (range 33-89). Most cases (594/613, 96.9%) were complete bilateral two-view (medio-lateral oblique and cranio-caudal) exams while a small subset (19/613 cases, 3.1%) was lacking one or two projections.

*Visual assessment of mammographic density*

Visual assessment of BD was independently performed through a dedicated web-based application by 21 radiologists, using the BI-RADS 4-class score. Exams from different vendors were mixed and presented in random order to avoid bias in comparing vendors; case order was the same for all independent readers.

On average, radiologists had 18-years experience (SD 8, range 5-27 years) of film-screen and digital mammography interpretation. Considering digital mammography alone, 20/21 readers had at least 3-years experience (mean 6, SD 3), had interpreted a mean of 8,442 digital mammograms in the year prior to the study (SD 6,730), and for 15/21 radiologists, screening exams accounted at least 50% of their readings. Readers routinely read mammograms by GE (13 of 21 readers, 62%), IMS (38%), Fuji (33%), Hologic (33%), Sectra (24%), and Siemens (14%); 13/21 readers (62%) routinely read images from multiple vendors.

*Reference standard*

The panel majority report (PMR), that is the mode of individual readings, was used as reference standard to compare visual and automated classification. Classes with equal counts (ties) were observed in 15/613 cases (2.4%). Ties could occur even with an odd number of readers; for instance, when ten readers classified an exam as class *b*, ten as class *c* and one as class *d*. Ties were resolved by randomly selecting among the two majority classes.

*Inter-observer agreement*

Inter-observer agreement was assessed by calculating Cohen's κ statistics for each reader pair, and for each reader with respect to PMR; overall panel agreement was assessed by Fleiss κ index. κ values from 0.00 to 0.20, from 0.21 to 0.40, from 0.41 to 0.60, from 0.61 to 0.80, and from 0.81 to 1.00 were interpreted as minimal, fair, moderate, substantial, and almost perfect agreement, respectively [14].

Agreement was assessed for binary classification (classes *c* and *d* collapsed as "dense" versus classes *a* and *b* collapsed as "non-dense") as well as for the 4 classes (*a*, *b*, *c*, and *d*) separately. Linear weighted κ was used for 4-class comparison.

*Automated breast density assessment*

The automated BD evaluator (ABDE) used in this study (*QUID*, prototype version, im3D SpA, Torino, Italy) automatically estimates BD by calculating the ratio of fibroglandular tissue area with respect to the total breast area on each view. The percentage value is then translated to a BI-RADS class (*a*, *b*, *c*, *d*) by applying a set of thresholds, calculated on a separate training dataset; the training set included images from the same vendors, was assessed by the same radiologist panel, and had similar density distribution compared to the present testing set. None of the cases used in this study was employed to train the ABDE algorithms.. Examples of ABDE results are provided in Figure 2.

For each exam, the automated BI-RADS class was obtained by taking the majority class among the four views; ties were resolved by random selection. Differences in ABDE within each patient were assessed by calculating the frequency of cases in which all four views (RCC, RMLO, LCC, LMLO) were classified in the same class, the frequency of cases in which one, two or three views were classified differently, and the maximum difference in the assigned BI-RADS categories.

Agreement of *QUID* ABDE with each reader and with the PMR was calculated using linear weighted κ statistics. In addition, classification accuracy was calculated as the ratio of the exams correctly classified by the automated system, with the panel as a reference, overall and separately for each density class. Agreement and accuracy were calculated for both binary and 4-class classification.

To account for the variability due to random class assignments, a simulation experiment was performed with ten repetitions. It is worth to notice all scores remain unchanged across repetitions, except in the case of ties. Mean and SD across repetitions were calculated for all agreement and accuracy measures.

*Analysis by vendor*

Readers agreement and ABDE performances were stratified by vendor. For each vendor subset, inter-observer agreement was assessed by averaging the κ values for all possible reader pairs to provide a single index of agreement; the mean κ value between each reader and the PMR was also calculated. To assess the effect of different radiologists' experience with each of the six vendor systems, inter-observer agreement for experienced vs. inexperienced readers was compared, where for each vendor the experienced group included only radiologists who routinely read images from that specific vendor.

Linear regression analysis was used to compare agreement among different vendors systems and among readers with different experience.

The agreement between the ABDE and the PMR, as well as ABDE classification accuracy, were separately calculated on each vendor subset; again, a simulation experiment was performed with ten repetitions, as detailed in the previous section.

**Results**

*Inter-reader agreement*

The dataset distribution of BD according to the PMR, which served as a reference standard, is shown in Figure 3.

Figure 4 shows the distribution of pairwise reader agreement for binary classification, including all 210 possible reader pair combinations (κ mean value 0.620, SD 0.137), while the agreement between each individual reader and PMR (for both binary and 4-class classification) is reported in Table 1. The agreement between each individual reader and PMR for binary classification was moderate for 4/21 readers, substantial for 9/21, and almost perfect for 8/21 (mean κ 0.736, SD 0.117, range 0.483–0.885).

Overall agreement for the reader panel (Fleiss κ index) was 0.602 (95% CI: 0.600-0.603) for binary classification and 0.400 (95% CI: 0.399-0.401) for 4-class classification.

*Agreement with the automated system*

The dataset distribution of BD according to *QUID* ABDE is reported in Figure 3. The agreement between each reader and *QUID* ABDE (for both binary and 4-class classification) is presented in Table 1. The agreement was moderate for 4/21 and substantial for 17/21 readers, with a mean κ value of 0.674 (SD 0.095, range 0.492– 0.779).

The agreement between *QUID* ABDE and PMR was almost perfect, with a mean κ value of 0.831 (SD 0.006), for binary classification and substantial, with a mean κ value of 0.699 (SD 0.006), for the 4-class classification.

*ABDE accuracy*

Ties resolved by random selection occurred in 97/613 cases (16%) for 4-class classification and in 28/613 cases (5%) for binary classification. Cases where *QUID* ABDE was uncertain between class *c* and class *d*, or class *b* and class *a*, would not count as ties as far as binary classification is concerned. In 252/613 cases (41.1%) all projections were classified in the same class, while in 238/613 (39.8%), 122/613 (19.9%) and 1/613 (0.2%) cases one, two or three projections respectively were scored differently. Only in 28/613 cases (4.6%), two projections of the same case differed by more than one class (e.g., RCC=3, RMLO=3, LCC=3, LMLO=1) according to ABDE.

Table 2 shows classification accuracy (mean and SD, after performing ten independent repetitions, as detailed in Methods) for each class and overall, for binary and 4-class classification. On average, *QUID* ABDE correctly classified 330/361 (91.3%) of *a-b* cases, 233/252 (92.5%) of *c-d* cases, and 563/613 (91.8%) of all cases, as compared to the PMR. Taking into account the variability due to ties, the overall binary accuracy ranged from 552/613 (90.1%) to 578/613 (94.3%), depending on all ties being correctly or incorrectly classified.

*By-vendor analysis*

Mean κ values for each reader pair and for each reader with respect to PMR, stratified by vendor, are reported in Table 3, while vendor-specific *QUID* ABDE's agreement with PMR and accuracy are reported in Table 4. Figure 5 shows the density distribution according to PMR for each vendor subset. Table 7 (on supplemental materials) shows the mean κ value for pairwise inter-observer agreement for experienced vs. inexperienced readers. Overall, the average pairwise agreement was moderate for IMS (mean κ: 0.59; 95% CI:0.55-0.61) and Sectra (mean κ:0.52; 95% CI:0.49-0.55), and substantial for other vendors. The number of experienced readers for each vendor ranged from 3 (Siemens) to 14 (GE). When only ratings by experienced readers for each vendor subset were included in the analysis, the average pairwise agreement across all vendors was substantial (mean κ: 0.67; 95% CI: 0.65-0.69). In contrast, a moderate agreement was observed in the group of inexperienced readers (mean κ: 0.60; 95% CI: 0.59-0.62; P<0.001). Vendor-specific agreement for inexperienced readers was moderate for IMS (mean κ: 0.53; 95% CI: 0.49-0.56) and Sectra (mean κ: 0.50; 95% CI: 0.47-0.53), and substantial for all other vendors. For experienced readers, κ values ranged from 0.63 (Hologic) to 0.74 (Sectra), all corresponding to substantial agreement (Table 3). Differences in agreement between experienced and inexperienced readers were statistically significant for Sectra (0.74 vs. 0.50; P<0.001), IMS (070 vs. 0,53; P<0.001) and FUJI (0.70 vs. 060; P=0.006), but not for Siemens (P=0.913) and Hologic (P=0.652). Differences for GE were statistically significant (0.77 vs. 0.65, P<0.001), but the agreement was substantial for both readers groups.

**Discussion**

In this study, an ABDE was tested on a multivendor set of digital mammograms, and the results, discretized in terms of BI-RADS classes, were compared with the BI-RADS classes provided by a large panel of experienced breast radiologists.

In principle, ABDE allows for assessing the percentage of fibroglandular tissue on a continuous scale, which is potentially a more accurate, precise and reproducible estimate if compared to visual assessment on a four point scale. However, a reference standard to evaluate ABDE does not exist. In the absence of a gold standard, the best estimate of the true measure is some combination (such as the mode or mean) of the raters' answers, which becomes the reference standard; in our case, the PMR was assumed as reference standard. This is the logical approach of the Bland-Altman method, where two measures are compared by plotting the difference between them to the reference standard given by the mean between the two [15]. This approach was followed in the specific field of BD evaluation by Ciatto et al [11]. Moreover, our knowledge about the clinical role of BD is mainly based on BI-RADS visual evaluation. As a consequence, correlating computed BD assessment with visual classification is a matter of interest for the medical community, with a potential for the use of an ABDE in clinical practice.

As a first step, we analyzed inter-rater agreement, to provide a benchmark for ABDE assessment and compare with results by previous studies. The mean κ for all possible readers combinations was 0.620 (SD 0.137), with a range of 0.210-0.842 (see Figure 4), confirming, on a larger multivendor dataset of digital images, the results reported by previous studies. A few reader pairs showed very low agreement (κ <0.3); however, those readers also had the lowest agreement with PMR, either overestimating or underestimating BD compared to the majority of readers. In previous studies, in 4-class classification, Redondo et al [12] found moderate agreement using unweighted κ (0.44) and substantial agreement using weighted κ (0.73) on a dataset of 100 film-screen mammograms read by twenty-one radiologists. Ciatto et al [11] observed a substantial agreement both for binary (mean κ 0.78, SD 0.06) and 4-class classification (mean κ 0.79, SD 0.05) among eleven breast radiologists reading 418 Hologic digital mammograms. Similar results were found by Bernardi et al [13].

The main finding of our study is that ABDE allowed for assigning the correct density BI-RADS binary classification (*a-b* vs. *c-d*) in almost 92% of the mammograms as compared to PMR used as a reference standard, giving an almost perfect agreement with a mean κ of 0.831. The average agreement of individual readers with the ABDE is slightly lower than their average agreement with PMR (0.674 vs. 0.736), but is comparable to the average readers pairwise agreement (0.620). Automated classification thus lies within the range of inter-rater variability observed in the present study, while offering several advantages as being completely automated and reproducible. *QUID* ABDE is indeed reproducible because the algorithms used to compute the percentage of fibroglandular tissue are completely deterministic.

These results favorably compare with studies previously reported for other ABDEs. Ciatto et al. [11] (11 radiologists, 418 exams, one vendor) found accuracies of 89% and 90% for class *a-b* and *c-d*, respectively, while *QUID* classified correctly over than 91% of the cases in both categories, on a larger and multi-vendor dataset. Mi Gweon et al. [16], in a study including 3 radiologists evaluating 778 exams from two vendors, found moderate agreement (κ=0.54) for 4-category classification, compared to a substantial agreement (κ=0.70) observed here with a larger panel.

ABDE BI-RADS class was separately estimated on each of the four mammographic views, and than the majority class was considered, so that missing or misclassified projections had a lesser impact on the overall classification. In around 5% of the cases (binary classification) ties happened, for instance in with the case of asymmetries between the left and right breasts, or when density values are very close to threshold values, and hence may "fall" on either side.

Notably, images from different vendors may present a variety of largely different "looks". Indeed, there were differences in the average reader agreement with PMR, but agreement was nevertheless substantial for all vendors; the average inter-observer pairwise agreement was moderate for IMS and Sectra, and substantial for all other vendors. Not all readers involved in this study were accustomed to read exams from so many different vendors, hence they were stratified according to the vendor (or vendors) they were most experienced with. Overall, the agreement with PMR improved when considering only experienced readers, especially for IMS and Sectra. In most cases, differences in agreement between experienced and inexperienced readers were statistically significant, but agreement was substantial for both groups for all vendors except IMS and Sectra. In our study, readers appeared to achieve higher agreement on images from the most common vendors (such as GE, Siemens and Hologic). For all vendors, the ABDE achieved a substantial or almost perfect agreement with PMR and an overall accuracy over 88%. Of note, the ABDE accuracy and agreement with PMR decreased along with readers' agreement with PMR: with increased reader variability, the system might be less capable of reproducing the "majority" reader.

This study has limitations. First, dataset BD was not uniformly distributed among the four classes and despite the relatively low average women age (55 years), class *d* was under-represented (only 8.8%). This could affect agreement assessment, since κ statistics is influenced by class prevalence [17]. Nevertheless, the number of extremely dense cases is similar to prior works; in Ciatto et al. [11], class *d* cases were only 6.2%. On the other hand, the study dataset is not representative of the BD distribution within the general screening population, depending also on the age at which screening is started and ended. In this study, both κ statistics and accuracy, using the PMR as a reference standard, were calculated. Notably, accuracy results may suggest higher ABDE performance, compared to the agreement analysis, as the accuracy measure (that is the ratio of the exams correctly classified by the automated system, with the panel as a reference) is not corrected for concordance based on chance. On the other hand, accuracy measures are easier to interpret than κ statistics and thus can be of practical value, when coupled with more complex statistics. By including accuracy values, we were also able to compare our results with previous studies, such as the one published by Ciatto et al [11].

Second, the ABDE was trained and tested on the same images radiologists read in their clinical routine (i.e. *for-presentation*), whose characteristics in terms of contrast and intensity distribution can widely vary across vendors and even mammographic units. However, the use of for-presentation instead of raw (i.e. *for-processing*) images, has also some advantages, as it is more easily integrated in clinical practice, and allows to retrospectively process available datasets, for which raw images are often not available. In any case, the ABDE showed comparable performance across vendors.

Third, thresholds used to convert the ABDE continuous values to the discrete BI-RADS classes were determined using an independent training dataset, however assessed by the same reader panel involved in this study. In principle, our results could not be perfectly generalizable to other readers; however, we sought to reduce this bias by including a large number of readers in the panel.

Finally, the use of two-dimensional mammographic images could not be the optimal choice for assessing the ratio of fibroglandular tissue in the breast three-dimensional volume. The use of tomosynthesis images for BD evaluation could possibly bring further improvements and more precise and reliable ABDE estimates [18].

In conclusion, the results show that *QUID* ABDE estimates are in good agreement with the majority report of a large panel of expert breast radiologists, as well as with the majority of individual readers. The observed performances suggest that the system is a viable alternative to visual classification and could be used as an automatic reproducible tool in tailored screening scenarios. Further studies are needed to validate the use of computed BD assessment in screening programs, and to understand how computed continuous BD measurements correlate with individual risk for cancer and sensitivity of mammography.

**Acknowledgement**

The authors would like to thank all the centers and the professionals who provided mammography exams for this study: in alphabetical order, APSS (Trento), A.O.U. Città della Salute e della Scienza (Turin), IRCCS Policlinico San Donato (Milan), ISPO (Florence), Ospedale Maggiore della Carità (Novara), Ospedale Regionale (Bolzano), Ospedale S.Andrea (Vercelli), and San Giovanni Bosco ASLTO2Nord (Turin).

The authors would like to acknowledge Dr. Stefano Ciatto for the helpful discussions and the suggestions he provided for this study and, in general, to the im3D research team: even after his passing, his teachings continue to drive breast imaging scientific research.

**Figures and tables**

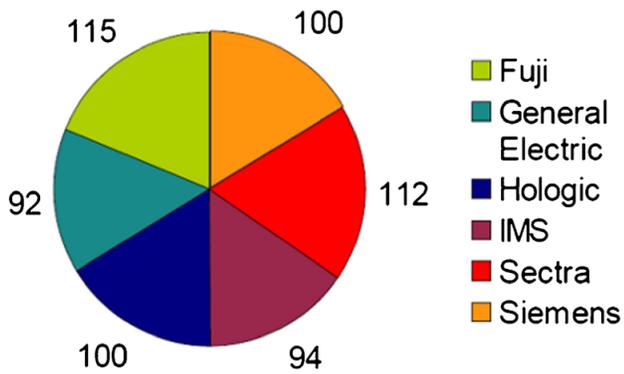

**Fig.1** Distribution (number of images per vendor) of the testing dataset

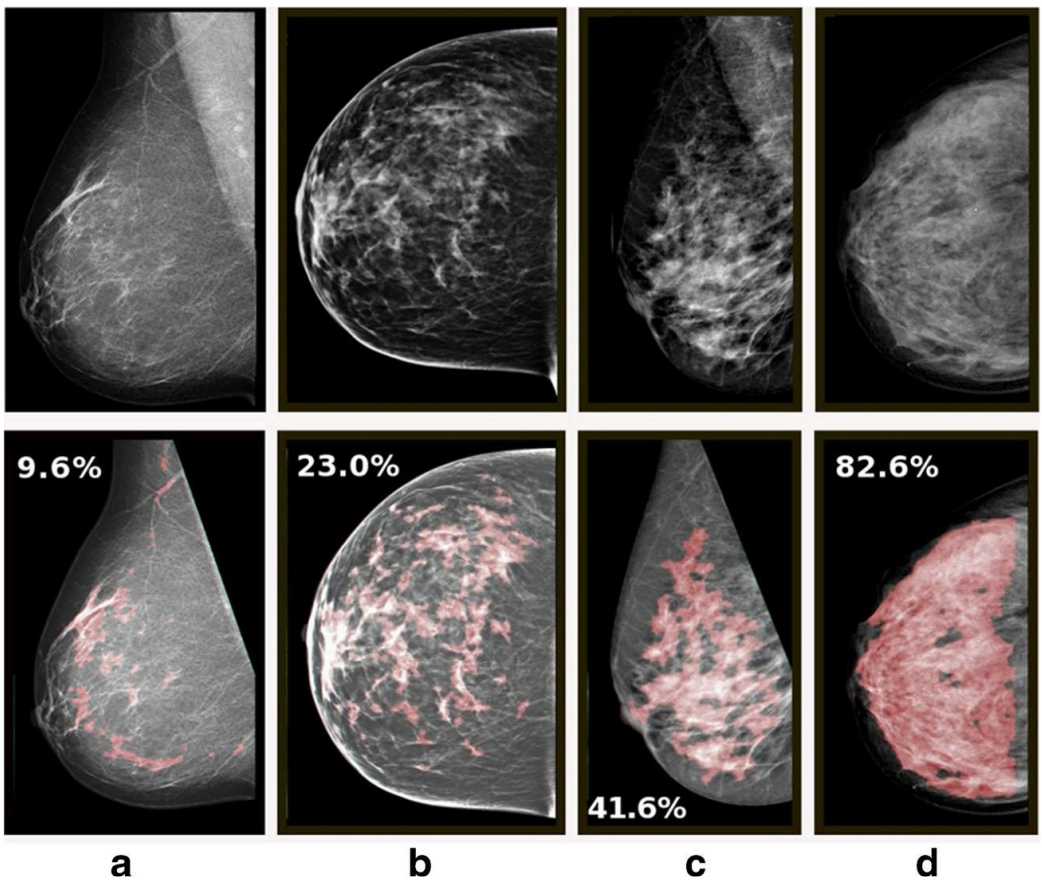

**Fig.2** Examples of mammograms for the four BI-RADS classes: (A) class *a*, FujiFilm Corporation, (B) class *b*, Hologic Inc., (C) class *c*, General Electric, (D) class *d*, Siemens. The segmentation of fibroglandular tissue by the automatic system is superimposed in red to the mammogram, and the calculated density value (in percentage) is shown for each mammogram.

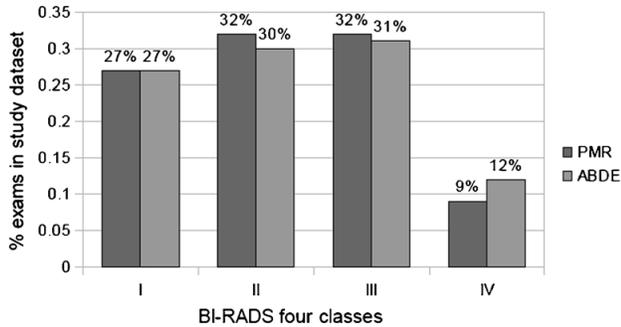

**Fig.3** Dataset density distribution according to the panel majority report (PMR) and the automated

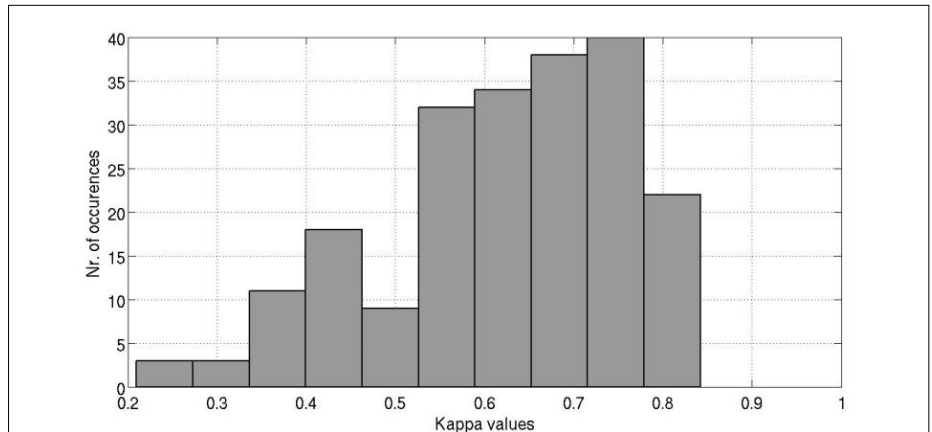

breast density evaluator (ABDE).

**Fig.4** Distribution of the pairwise reader agreement for the binary classification (*a-b* vs. *c-d*).

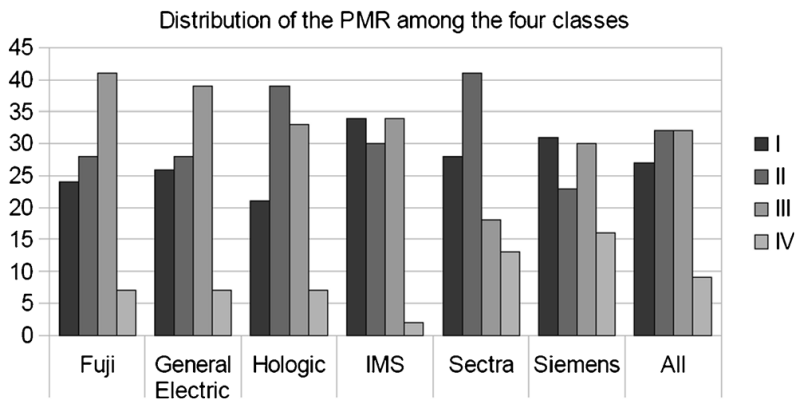

**Fig.5** Density distribution according to the panel majority report (PMR), for each vendor subset and for the whole dataset.

| Reader | PMR | | ABDE | |
| --- | --- | --- | --- | --- |
| | Binary classification | 4-class classification | Binary classification | 4-class classification |
| R1 | 0.757 | 0.704 | 0.731 | 0.642 |
| R2 | 0.564 | 0.536 | 0.512 | 0.462 |
| R3 | 0.687 | 0.657 | 0.663 | 0.580 |
| R4 | 0.554 | 0.513 | 0.538 | 0.441 |
| R5 | 0.813 | 0.656 | 0.770 | 0.564 |
| R6 | 0.822 | 0.834 | 0.761 | 0.668 |
| R7 | 0.802 | 0.766 | 0.694 | 0.609 |
| R8 | 0.820 | 0.841 | 0.746 | 0.668 |
| R9 | 0.792 | 0.730 | 0.722 | 0.621 |
| R10 | 0.885 | 0.797 | 0.778 | 0.661 |
| R11 | 0.684 | 0.726 | 0.621 | 0.574 |
| R12 | 0.772 | 0.638 | 0.704 | 0.515 |
| R13 | 0.539 | 0.632 | 0.484 | 0.504 |
| R14 | 0.838 | 0.799 | 0.743 | 0.641 |
| R15 | 0.828 | 0.831 | 0.733 | 0.613 |
| R16 | 0.701 | 0.662 | 0.668 | 0.575 |
| R17 | 0.796 | 0.784 | 0.694 | 0.625 |
| R18 | 0.483 | 0.486 | 0.498 | 0.421 |
| R19 | 0.659 | 0.579 | 0.604 | 0.502 |
| R20 | 0.880 | 0.842 | 0.750 | 0.661 |
| R21 | 0.781 | 0.760 | 0.748 | 0.650 |
| Mean | 0.736 | 0.703 | 0.674 | 0.581 |
| SD | 0.117 | 0.111 | 0.095 | 0.078 |

**Table 1** Agreement (κ value) of each individual reader (R1 to R21) with the panel majority report (PMR), and with the automated breast density evaluator (ABDE), for both binary and 4-class classifications. Mean and standard deviation (SD) of the κ values of all readers are also reported.

|  | Mean % | SD |
|---|---|---|
| *Binary classification* | | |
| **Accuracy a-b** | 91.30 | 0.47 |
| **Accuracy c-d** | 92.46 | 0.53 |
| **Overall Accuracy** | 91.78 | 0.28 |
| *4-class classification* | | |
| **Accuracy a** | 76.11 | 1.21 |
| **Accuracy b** | 64.59 | 1.54 |
| **Accuracy c** | 68.69 | 1.39 |
| **Accuracy d** | 56.30 | 2.34 |
| **Overall Accuracy** | 68.32 | 0.80 |

**Table 2** Accuracy of *QUID* automated breast density evaluator classification (number of exams correctly classified divided by the total number of cases), for each BI-RADS density class (*a* to *d*) and overall, for the binary and 4-class classification.

| Vendor | Average age | Number of readers with experience with each vendor | κ for the whole 21-radiologist panel | | | | κ by readers with experience with each vendor | | | |
|---|---|---|---|---|---|---|---|---|---|---|
| | | | Readers vs PMR | | Pairwise reader agreement | | Readers vs PMR | | Pairwise reader agreement | |
| | | | Mean | SD | Mean | SD | Mean | SD | Mean | SD |
| **General Electric** | 56 | 14 | 0.80 | 0.11 | 0.69 | 0.13 | 0.77 | 0.12 | 0.65 | 0.14 |
| **Siemens** | 56 | 3 | 0.78 | 0.11 | 0.68 | 0.14 | 0.70 | 0.16 | 0.69 | 0.14 |
| **Hologic** | 55 | 7 | 0.74 | 0.13 | 0.62 | 0.15 | 0.73 | 0.13 | 0.63 | 0.14 |
| **Fuji** | 52 | 7 | 0.73 | 0.13 | 0.61 | 0.15 | 0.78 | 0.13 | 0.70 | 0.08 |
| **IMS** | 57 | 8 | 0.71 | 0.14 | 0.59 | 0.16 | 0.81 | 0.10 | 0.70 | 0.11 |
| **Sectra** | 57 | 5 | 0.65 | 0.18 | 0.52 | 0.18 | 0.72 | 0.11 | 0.74 | 0.05 |
| **All vendors** | 55 | 21 | 0.74 | 0.12 | 0.62 | 0.14 | 0.77 | 0.11 | 0.67 | 0.13 |

**Table 3** Analysis by vendor: average κ values for each reader pair (pairwise reader agreement) and for each reader with respect to panel majority report (readers vs PMR), for the whole panel of 21

radiologists and for experienced readers (i.e. readers with experience with each vendor). Mean and standard deviation (SD) of the κ values for all vendors are also reported.

| Vendor | ABDE vs PMR κ | | ABDE Accuracy | |
|---|---|---|---|---|
| | Mean | SD | Mean | SD |
| **Siemens** | 0.90 | 0.01 | 95.0% | 0.67 |
| **General Electric** | 0.87 | 0.02 | 93.4% | 1.08 |
| **Hologic** | 0.82 | 0.02 | 91.2% | 0.92 |
| **Fuji** | 0.85 | 0.03 | 92.7% | 1.65 |
| **IMS** | 0.79 | 0.01 | 89.7% | 0.72 |
| **Sectra** | 0.74 | 0.02 | 88.9% | 0.86 |

**Table 4** Vendor-specific agreement of *QUID* automatic breast density evaluator (ABDE) versus panel majority report (PMR) as measured by κ values and ABDE classification accuracy. Mean and standard deviation (SD) were calculated on 10 separate simulation experiment repetitions.

**Electronic Supplementary Material**

| | R1 | R2 | R3 | R4 | R5 | R6 | R7 | R8 | R9 | R10 | R11 | R12 | R13 | R14 | R15 | R16 | R17 | R18 | R19 | R20 | R21 | PMR Bin. | PMR 4class | ABDE Bin. | ABDE 4class |
|---|---|---|---|---|---|---|---|---|---|---|---|---|---|---|---|---|---|---|---|---|---|---|---|---|---|
| R1 | - | 0.406 | 0.761 | 0.637 | 0.721 | 0.660 | 0.606 | 0.629 | 0.597 | 0.725 | 0.512 | 0.762 | 0.386 | 0.740 | 0.648 | 0.763 | 0.619 | 0.632 | 0.590 | 0.677 | 0.720 | **0.757** | **0.704** | **0.731** | **0.642** |
| R2 | 0.406 | - | 0.343 | 0.278 | 0.466 | 0.589 | 0.650 | 0.673 | 0.663 | 0.570 | 0.740 | 0.421 | 0.737 | 0.490 | 0.619 | 0.357 | 0.645 | 0.222 | 0.411 | 0.627 | 0.432 | **0.564** | **0.536** | **0.512** | **0.462** |
| R3 | 0.761 | 0.343 | - | 0.676 | 0.693 | 0.589 | 0.553 | 0.549 | 0.526 | 0.658 | 0.431 | 0.758 | 0.319 | 0.726 | 0.559 | 0.802 | 0.539 | 0.709 | 0.557 | 0.606 | 0.697 | **0.687** | **0.657** | **0.663** | **0.580** |
| R4 | 0.637 | 0.278 | 0.676 | - | 0.581 | 0.477 | 0.438 | 0.453 | 0.431 | 0.522 | 0.348 | 0.651 | 0.258 | 0.601 | 0.474 | 0.682 | 0.431 | 0.665 | 0.459 | 0.494 | 0.616 | **0.554** | **0.513** | **0.538** | **0.441** |
| R5 | 0.721 | 0.466 | 0.693 | 0.581 | - | 0.759 | 0.669 | 0.705 | 0.659 | 0.747 | 0.567 | 0.730 | 0.440 | 0.726 | 0.699 | 0.695 | 0.649 | 0.548 | 0.563 | 0.716 | 0.740 | **0.813** | **0.656** | **0.770** | **0.564** |
| R6 | 0.660 | 0.589 | 0.589 | 0.477 | 0.759 | - | 0.755 | 0.808 | 0.729 | 0.767 | 0.687 | 0.681 | 0.593 | 0.697 | 0.768 | 0.602 | 0.706 | 0.424 | 0.596 | 0.787 | 0.689 | **0.822** | **0.834** | **0.761** | **0.668** |
| R7 | 0.606 | 0.650 | 0.553 | 0.438 | 0.669 | 0.755 | - | 0.777 | 0.779 | 0.798 | 0.731 | 0.638 | 0.633 | 0.685 | 0.780 | 0.549 | 0.752 | 0.373 | 0.560 | 0.807 | 0.659 | **0.802** | **0.766** | **0.694** | **0.609** |
| R8 | 0.629 | 0.673 | 0.549 | 0.453 | 0.705 | 0.808 | 0.777 | - | 0.795 | 0.791 | 0.790 | 0.655 | 0.652 | 0.708 | 0.842 | 0.574 | 0.764 | 0.380 | 0.604 | 0.818 | 0.650 | **0.820** | **0.841** | **0.746** | **0.668** |
| R9 | 0.597 | 0.663 | 0.526 | 0.431 | 0.659 | 0.729 | 0.779 | 0.795 | - | 0.763 | 0.760 | 0.603 | 0.687 | 0.675 | 0.805 | 0.538 | 0.810 | 0.362 | 0.583 | 0.788 | 0.663 | **0.792** | **0.730** | **0.722** | **0.621** |
| R10 | 0.725 | 0.570 | 0.658 | 0.522 | 0.747 | 0.767 | 0.798 | 0.791 | 0.763 | - | 0.699 | 0.766 | 0.562 | 0.785 | 0.814 | 0.684 | 0.757 | 0.459 | 0.600 | 0.832 | 0.690 | **0.885** | **0.797** | **0.778** | **0.661** |
| R11 | 0.512 | 0.740 | 0.431 | 0.348 | 0.567 | 0.687 | 0.731 | 0.790 | 0.760 | 0.699 | - | 0.529 | 0.721 | 0.587 | 0.757 | 0.452 | 0.718 | 0.288 | 0.495 | 0.770 | 0.535 | **0.684** | **0.726** | **0.621** | **0.574** |
| R12 | 0.762 | 0.421 | 0.758 | 0.651 | 0.730 | 0.681 | 0.638 | 0.655 | 0.603 | 0.766 | 0.529 | - | 0.400 | 0.742 | 0.668 | 0.792 | 0.612 | 0.618 | 0.573 | 0.684 | 0.716 | **0.772** | **0.638** | **0.704** | **0.515** |
| R13 | 0.386 | 0.737 | 0.319 | 0.258 | 0.440 | 0.593 | 0.633 | 0.652 | 0.687 | 0.562 | 0.721 | 0.400 | - | 0.467 | 0.607 | 0.347 | 0.629 | 0.210 | 0.397 | 0.600 | 0.417 | **0.539** | **0.632** | **0.484** | **0.504** |
| R14 | 0.740 | 0.490 | 0.726 | 0.601 | 0.726 | 0.697 | 0.685 | 0.708 | 0.675 | 0.785 | 0.587 | 0.742 | 0.467 | - | 0.729 | 0.727 | 0.718 | 0.551 | 0.588 | 0.766 | 0.752 | **0.838** | **0.799** | **0.743** | **0.641** |
| R15 | 0.648 | 0.619 | 0.559 | 0.474 | 0.699 | 0.768 | 0.780 | 0.842 | 0.805 | 0.814 | 0.757 | 0.668 | 0.607 | 0.729 | - | 0.585 | 0.803 | 0.401 | 0.583 | 0.806 | 0.683 | **0.828** | **0.831** | **0.733** | **0.613** |
| R16 | 0.763 | 0.357 | 0.802 | 0.682 | 0.695 | 0.602 | 0.549 | 0.574 | 0.538 | 0.684 | 0.452 | 0.792 | 0.347 | 0.727 | 0.585 | - | 0.555 | 0.705 | 0.539 | 0.625 | 0.679 | **0.701** | **0.662** | **0.668** | **0.575** |
| R17 | 0.619 | 0.645 | 0.539 | 0.431 | 0.649 | 0.706 | 0.752 | 0.764 | 0.810 | 0.757 | 0.718 | 0.612 | 0.629 | 0.718 | 0.803 | 0.555 | - | 0.382 | 0.547 | 0.786 | 0.686 | **0.796** | **0.784** | **0.694** | **0.625** |
| R18 | 0.632 | 0.222 | 0.709 | 0.665 | 0.548 | 0.424 | 0.373 | 0.380 | 0.362 | 0.459 | 0.288 | 0.618 | 0.210 | 0.551 | 0.401 | 0.705 | 0.382 | - | 0.394 | 0.418 | 0.578 | **0.483** | **0.486** | **0.498** | **0.421** |
| R19 | 0.590 | 0.411 | 0.557 | 0.459 | 0.563 | 0.596 | 0.560 | 0.604 | 0.583 | 0.600 | 0.495 | 0.573 | 0.397 | 0.588 | 0.583 | 0.539 | 0.547 | 0.394 | - | 0.578 | 0.587 | **0.659** | **0.579** | **0.604** | **0.502** |
| R20 | 0.677 | 0.627 | 0.606 | 0.494 | 0.716 | 0.787 | 0.807 | 0.818 | 0.788 | 0.832 | 0.770 | 0.684 | 0.600 | 0.766 | 0.806 | 0.625 | 0.786 | 0.418 | 0.578 | - | 0.699 | **0.880** | **0.842** | **0.750** | **0.661** |
| R21 | 0.720 | 0.432 | 0.697 | 0.616 | 0.740 | 0.689 | 0.659 | 0.650 | 0.663 | 0.690 | 0.535 | 0.716 | 0.417 | 0.752 | 0.683 | 0.679 | 0.686 | 0.578 | 0.587 | 0.699 | - | **0.781** | **0.760** | **0.748** | **0.650** |
| *Mean* | *0.640* | *0.517* | *0.603* | *0.509* | *0.654* | *0.668* | *0.660* | *0.681* | *0.661* | *0.699* | *0.606* | *0.650* | *0.503* | *0.673* | *0.681* | *0.613* | *0.655* | *0.466* | *0.540* | *0.694* | *0.644* | ***0.736*** | ***0.703*** | ***0.674*** | ***0.581*** |
| *SD* | *0.105* | *0.150* | *0.130* | *0.125* | *0.092* | *0.100* | *0.117* | *0.122* | *0.124* | *0.103* | *0.148* | *0.105* | *0.154* | *0.090* | *0.119* | *0.126* | *0.116* | *0.148* | *0.068* | *0.112* | *0.090* | ***0.114*** | ***0.108*** | ***0.095*** | ***0.078*** |

**Table 5** Inter-reader agreement analysis: pairwise inter-reader agreement, considering all possible combinations of 21 readers, are reported, along with agreement of each reader with the panel majority report (PMR) and with the automated breast density evaluator (ABDE). For the ABDE and the PMR both binary and 4-class classifications were taken into account, while for reason of simplicity pairwise inter-rater agreement is reported for binary classification only. Finally, mean and standard deviation of individual κ values for the reader panel are included.

|      | *a*  | *b*  | *c*  | *d*  |
|------|------|------|------|------|
| R3   | 15%  | 29%  | 41%  | 15%  |
| R4   | 11%  | 30%  | 43%  | 16%  |
| R5   | 8%   | 44%  | 36%  | 11%  |
| R6   | 29%  | 33%  | 24%  | 13%  |
| R7   | 40%  | 26%  | 26%  | 8%   |
| R8   | 32%  | 34%  | 26%  | 8%   |
| R9   | 38%  | 29%  | 28%  | 4%   |
| R10  | 34%  | 26%  | 21%  | 19%  |
| R11  | 30%  | 43%  | 25%  | 2%   |
| R12  | 12%  | 38%  | 48%  | 2%   |
| R13  | 35%  | 44%  | 18%  | 2%   |
| R14  | 23%  | 31%  | 30%  | 15%  |
| R15  | 28%  | 36%  | 30%  | 6%   |
| R16  | 14%  | 30%  | 40%  | 16%  |
| R17  | 26%  | 40%  | 25%  | 9%   |
| R18  | 8%   | 23%  | 60%  | 8%   |
| R19  | 30%  | 27%  | 26%  | 13%  |
| R20  | 34%  | 30%  | 25%  | 12%  |
| R21  | 21%  | 30%  | 33%  | 16%  |
| **PMR**  | **27%** | **32%** | **32%** | **9%**  |
| **ABDE** | **27%** | **30%** | **31%** | **12%** |

**Table 6** Distribution of reports by the 21 readers, the panel majority report (PMR) and the automated breast density evaluator (ABDE).

| Vendor | κ experienced readers group | | κ inexperienced readers group | | P-value |
| --- | --- | --- | --- | --- | --- |
| | Mean | 95% C.I. | Mean | 95% C.I. | |
| **General Electric** | 0.65 | (0.63-0.68) | 0.77 | (0.71-0.84) | <0.001 |
| **Siemens** | 0.69 | (0.55-0.84) | 0.70 | (0.67-0.72) | 0.913 |
| **Hologic** | 0.63 | (0.57-0.68) | 0.64 | (0.61-0.67) | 0.652 |
| **Fuji** | 0.70 | (0.65-0.75) | 0.60 | (0.57-0.63) | 0.006 |
| **IMS** | 0.70 | (0.66-0.75) | 0.53 | (0.49-0.56) | <0.001 |
| **Sectra** | 0.74 | (0.66-0.81) | 0.50 | (0.47-0.53) | <0.001 |

**Table 7** Analysis by vendor and experience: average κ values for each reader pair (pairwise reader agreement) in experienced vs. inexperienced readers. For each vendor, the experienced (inexperienced) group include readers who routinely read (do not read) images from that vendor.